\documentclass{article}%
\usepackage{amssymb}
\usepackage{amsmath}%
\usepackage{amsfonts}%
\usepackage{graphicx}
\newtheorem{theorem}{Theorem}

\newtheorem{definition}[theorem]{Definition}
\newtheorem{example}[theorem]{Example}

\newtheorem{proposition}[theorem]{Proposition}

\newenvironment{proof}[1][Proof]{\textbf{#1.} }{\ \rule{0.5em}{0.5em}}
\begin{document}

\title{HOW\ CAN\ WE\ OBSERVE\ AND\ DESCRIBE\ CHAOS?\ \ \\\ \ }
\author{Andrzej Kossakowski$\dagger$, Masanori Ohya$\ddagger$ and Yosio
Togawa$\ddagger$\\$\dagger$Institute of Physics\\N. Copernicus University, Grudziadzka 5, 87-100 Torun, Poland\\$\ddagger$Department of Information Sciences\\Tokyo University of Science, Noda City, Chiba 278-8510, Japan}
\maketitle

\section{Introduction}

There exist several trials to describe chaos appeared in classical or quantum
dynamical systems \cite{AOW-MO}. One of the present authors introduced
Information Dynamics(ID for short) \cite{O2} as a frame to discuss complexity
and chaos appeared in various fields, in which he tried to find a common basis
by synthesizing the state change (dynamics) and the complexity associated with
dynamical systems. Since then, ID has been applied to several different topics
\cite{IKO.O3}, among which a chaos degree, a quantity measuring the degree of
chaos associated with a dynamics, was introduced by means of the complexities
in ID and its entropic version (called Entropic Chaos Degree (EDC for short))
has been computed numerically for rather famous chaotic dynamics such as
logistic map, baker's transformation, Tinkerbel map. It is surprised that the
result of the ECD exactly macthes to that of Lyapunov exponent in the case
that the later can be computed. Moreover the algorithm computing the ECD is
much easier than that of Lyapunov exponent, so that the ECD can be almost
always computable even when the Lyapunov exponent can not be so. However there
are some unclear points in both conceptually and mathematically why the ECD
could be so successful for computational experiments. In this paper we study
these points and propose a new description of chaos.

In Section 2, we briefly review Information Dynamics and Chaos Degree, and in
Section 3 the entropic chaos degree and its algorithm are recalled with a
computational result. In Section 4, a new way judging chaos from a given
dynamics is discussed based on the ECD, that is, we propose a new view to
define chaos of dynamical systems.

\section{Information Dyanamics and Chaos Degree}

We briefly review what ID is. Let $(\mathcal{A},\mathfrak{S},\alpha(G))$ be an
input (or initial) system and $(\overline{\mathcal{A}},\overline{\mathfrak{S}%
},\overline{\alpha}(\overline{G}))$ be an output (or final) system. Here
$\mathcal{A}$ is a set of some objects to be observed and $\mathfrak{S}$ is a
set of some means to get the observed value, $\alpha(G)$ describes a certain
evolution of system with a parameter $g$ in a certain set $G$. Often we have
$\mathcal{A}=\overline{\mathcal{A}}$, $\mathfrak{S}=\overline{\mathfrak{S}}$,
$\alpha=\overline{\alpha},$ $G=\overline{G}$. Therefore it can be said

\begin{center}
[Giving a mathematical structure to input and output triples

$\equiv$ Having a theory]
\end{center}

The dynamics of state change is described by a channel, that is, a map
$\Lambda^{\ast}$: $\mathfrak{S}\rightarrow\overline{\mathfrak{S}}$ (sometimes
$\mathfrak{S}\rightarrow\mathfrak{S}$). The fundamental point of ID is that ID
contains two complexities in itself. Let $(\mathcal{A}_{t},\mathfrak{S}%
_{t},\alpha^{t}(G^{t}))$ be the total system of $(\mathcal{A},\mathfrak{S}%
,\alpha)$ and $(\overline{\mathcal{A}},\overline{\mathfrak{S}},\overline
{\alpha})$, and $\mathcal{S}$ be a subset of $\mathfrak{S}$ in which we are
measuring observables (e.g., $\mathcal{S}$ is the set of all KMS or stationary
states in C*-system). Two complexities are denoted by $C$ and $T$. $C$ is the
complexity of a state $\varphi$ measured from a reference system $\mathcal{S}%
$, in which we actually observe the objects in $\mathcal{A}$ and $T$ is the
transmitted complexity associated with a state change $\varphi\rightarrow
\Lambda^{\ast}\varphi$, both of which should satisfy the following properties :

\noindent$\langle$\textbf{Aximos of complexities}$\rangle$

\begin{enumerate}
\item[(i)] For any $\varphi\in\mathcal{S}\subset\mathfrak{S}$,
\[
C^{\mathcal{S}}(\varphi)\geq0,\ T^{\mathcal{S}}(\varphi;\Lambda^{\ast})\geq0
\]

\item[(ii)] For any orthogonal\ bijection\ $j:ex\mathfrak{S}\rightarrow
ex\mathfrak{S}$, the set of all extremal points of $\mathfrak{S}$,
\[
C^{j(\mathcal{S})}(j(\varphi))=C^{\mathcal{S}}(\varphi)
\]
\[
T^{j(\mathcal{S})}(j(\varphi);\Lambda^{\ast})=T^{\mathcal{S}}(\varphi
;\Lambda^{\ast})
\]

\item[(iii)] For $\Phi\equiv\varphi\otimes\psi\in\mathcal{S}_{t}%
\subset\mathfrak{S}_{t}$,
\[
C^{\mathcal{S}_{t}}(\Phi)=C^{\mathcal{S}}(\varphi)+C^{\overline{\mathcal{S}}%
}(\psi)
\]

\item[(iv)] $0\le T^{\mathcal{S}}(\varphi;\Lambda^{*})\le C^{\mathcal{S}%
}(\varphi)$

\item[(v)] $T^{\mathcal{S}}(\varphi;id)=C^{\mathcal{S}}(\varphi)$, where
``$id$'' is an identity map from $\mathfrak{S}$ to $\mathfrak{S}$.
\end{enumerate}

Instead of (iii), when ``(iii') $\Phi\in\mathcal{S}_{t}\subset\mathfrak{S}%
_{t}$, put $\varphi\equiv\Phi\upharpoonright\mathcal{A}$ (i.e., the
restriction of $\Phi$ to $\mathcal{A}$), $\psi\equiv\Phi\upharpoonright
\overline{\mathcal{A}}$, $C^{\mathcal{S}_{t}}(\Phi)\leq C^{\mathcal{S}%
}(\varphi)+C^{\overline{\mathcal{S}}}(\psi)$ '' is satisfied, $C$ and $T$ is
called a pair of strong complexity. Therefore ID is defined as follows:

\begin{definition}
\emph{: }\ Information Dynamics is described by
\[%
\begin{array}
[c]{c}%
\left(  \mathcal{A},\mathfrak{S},\alpha(G);\overline{\mathcal{A}}%
,\overline{\mathfrak{S}},\overline{\alpha}(\overline{G});\Lambda^{\ast
};C^{\mathcal{S}}(\varphi),T^{\mathcal{S}}(\varphi;\Lambda^{\ast})\right) \\
\mathrm{and\ some\ relations}\ R\ \mathrm{among\ them}.
\end{array}
\]

\end{definition}

In the framework of ID, we have to

\begin{enumerate}
\item[(i)] mathematically determine
\[
\mathcal{A},\mathfrak{S},\alpha(G);\overline{\mathcal{A}},\overline
{\mathfrak{S}},\overline{\alpha}(\overline{G}),
\]

\item[(ii)] choose $\Lambda^{\ast}$ and $R$, and

\item[(iii)] define $C^{\mathcal{S}}(\varphi)$, $T^{\mathcal{S}}%
(\varphi;\Lambda^{\ast})$.
\end{enumerate}

In ID, several different topics can be treated on a common standing point so
that we can find a new clue bridging several fields.

We assume $\overline{\mathcal{A}}=\mathcal{A}$ for simlicity in the sequel.
For a certain subset $\mathcal{S}$ (called the reference space) of
$\mathfrak{S}$ and a state $\varphi\in\mathcal{S}$, there exists a
decomposition of the state $\varphi$ into a mixture of extreme (pure) states
such that%

\[
\varphi=\int_{\mathcal{S}}\omega d\mu
\]
This extremal decomposition of $\varphi$ describes the degree of mixture of
$\varphi$ in the reference space $\mathcal{S}$. The measure $\mu$ is not
always unique, so that the set of all such measures is denoted by $M_{\varphi
}\left(  \mathcal{S}\right)  .$

For instance, when $(\mathcal{A},\mathfrak{S})$ and is a C*-system containing
both classical and quantum systems; that is, $\mathcal{A}$ and is a C* algebra
and $\mathfrak{S}$ is the set of all states on $\mathcal{A}$, the reference
space $\mathcal{S}$ is a weak* compact convex subset of $\mathfrak{S}$ and the
measure $\mu$ is not uniquely determined unless $\mathcal{S}$ is the Schoque
simplex$.$ In this paper we will not go to the details of such general
mathematical discussion.

A measure of chaos produced by dynamics $\Lambda^{\ast}$ is defined in
\cite{O5,O6}:

\begin{definition}
(1)$\psi$ is more chaotic than $\varphi$ if $C(\psi)\geq C(\varphi)$.

(2)When $\varphi\in\mathcal{S}$ changes to $\Lambda^{\ast}\varphi$,
the\textit{\ chaos} degree associated to this state change (dynamics)
$\Lambda^{\ast}$ is given by
\[
D^{\mathcal{S}}\left(  \varphi;\Lambda^{\ast}\right)  =\inf\left\{
\int_{\mathcal{S}}C^{\mathcal{S}}\left(  \Lambda^{\ast}\omega\right)  d\mu
;\mu\in M_{\varphi}\left(  \mathcal{S}\right)  \right\}  .
\]

\end{definition}

\begin{definition}
A dynamics $\Lambda^{\ast}$ produces chaos iff $D^{\mathcal{S}}\left(
\varphi;\Lambda^{\ast}\right)  >0.$
\end{definition}

It is important to note here that the dynamics $\Lambda^{\ast}$ in the
definition is not necessarily same as original dynamics (channel) but is one
reduced from the original one such that it causes an evolution for a certain
observed value like orbit. However for simplicity we often use the same
notation in this paper. In some cases, the above chaos degree $D^{\mathcal{S}%
}\left(  \varphi;\Lambda^{\ast}\right)  $ can be expressed as%

\[
D^{\mathcal{S}}\left(  \varphi;\Lambda^{\ast}\right)  =C^{\mathcal{S}}\left(
\Lambda^{\ast}\varphi\right)  -T^{\mathcal{S}}(\varphi;\Lambda^{\ast}).
\]

\section{Entropic Chaos Degree and its Algorithm}

Although there exist several complexities \cite{O4}, one of the most useful
examples of $C$ and $T$ are Shannon's entropy and mutual entropy in classical
systems (von Neumann entropy and quantum mutual entropy in quantum systems
\cite{OP}), respectively.

The concept of entropy was introduced and developed to study the topics such
as irreversible behavior, symmetry breaking, amount of information
transmission, so that it originally describes a certain chaotic property of state.

\smallskip Let us recall the simplest case of $C$ and $T,$ that is, Shannon's
entropy and mutual entropy. In classical communication systems, an input state
$\varphi$ is a probability distribution $p=\left(  p_{k}\right)  =\sum
_{k}p_{k}\delta_{k}$ and a channel $\Lambda^{\ast}$ is a transition
probability $\left(  t_{i,j}\right)  ,$ so that the compound state of
$\varphi$ and its output $\overline{\varphi}$ ($\equiv$ $\overline{p}=\left(
\overline{p}_{i}\right)  =\Lambda^{\ast}p$) is the joint distribution
$r=\left(  r_{i,j}\right)  $ with $r_{i,j}\equiv t_{i,j}p_{j}.$ Then the
complexities $C$ and $T$ are given as%

\[%
\begin{array}
[c]{l}%
C\left(  p\right)  =S\left(  p\right)  =-\sum_{k}p_{k}\log p_{k},\\
T\left(  p;\Lambda^{\ast}\right)  =I\left(  p;\Lambda^{\ast}\right)
=\sum_{i,j}r_{i,j}\log\frac{r_{i,j}}{p_{j}\overline{p}_{i}}.
\end{array}
\]
Thus the entropic chaos degree of the channel $\Lambda^{\ast}$ becomes

\begin{definition}%
\[
D\left(  p;\Lambda^{\ast}\right)  =S\left(  \Lambda^{\ast}p\right)
-I(p;\Lambda^{\ast}).
\]

\end{definition}

Quantum version of the above entropic chaos degree was discussed in
\cite{IKO2,O6}, on which we will briefly review here in the case of usual
Hilbert space expression. Let $\rho$ be a quantum state, namely, a density
operator on a Hilbert space $\mathcal{H},$ and $\Lambda^{\ast}$ be a channel
sending the set $\mathfrak{S}$ of all states on $\mathcal{H}$ into itself.
Then the entropic chaos degree is defined by%

\[
D\left(  \rho;\Lambda^{\ast}\right)  =\inf\left\{  \sum_{k}\lambda_{k}S\left(
\Lambda^{\ast}E_{k}\right)  ;\left\{  E_{k}\right\}  \in\mathcal{E}\right\}  ,
\]
where $\mathcal{E}$ is the set of all Schatten decompositions (i.e., one
dimensional spectral decompositions) of the state $\rho:=$ $\sum_{k}%
\lambda_{k}E_{k},$ and $S$ is the von Neumann entropy.

\subsection{Algorithm Computing Chaos Degree}

In order to observe a chaos produced by a dynamics, one often looks at the
behaivor of orbits made by that dynamics, more generally, looks at the
behavior of a certain observed value. Therefore in our scheme we directly
compute the chaos degree once a dynamics is explicitly given as a state change
of a system. However even when the direct calculation does not show a chaos, a
chaos will appear if one forcuses to some aspect of the state change, e.g., a
certain observed value which may be called orbit as usual. The algorithm
computing the chaos degree for a dynamic is the following two cases
\cite{O5,O6,IOV1,IKO2}:

$\left(  1\right)  $\emph{\ Dynamics is given by }$\frac{dx}{dt}=\digamma
_{t}\left(  x\right)  $\emph{\ with }$x\in I\equiv\left[  a,b\right]
^{\mathbf{N}}\subset\mathbf{R}^{\mathbf{N}}\ :$ First find a difference
equation $x_{n+1}=\digamma\left(  x_{n}\right)  $ with a map $\digamma$ on
$I\equiv\left[  a,b\right]  ^{\mathbf{N}}\subset\mathbf{R}^{\mathbf{N}}$ into
itself, $\sec$ondly let $I\equiv\bigcup_{k}A_{k}$ be a finite partation with
$A_{i}\cap A_{j}=\emptyset$ $\left(  i\neq j\right)  .$ Then the state
$\varphi^{\left(  n\right)  }$ of the orbit determined by the difference
equation is defined by the probabilty distribution $\left(  p_{i}^{\left(
n\right)  }\right)  ,$ that is, $\varphi^{\left(  n\right)  }=\sum_{i}%
p_{i}^{\left(  n\right)  }\delta_{i},$ where for a given initial value $x\in
I$ and the characteristic function $1_{A}$%

\[
p_{i}^{\left(  n\right)  }\equiv\frac{1}{n+1}\sum_{k=m}^{m+n}1_{A_{i}}\left(
\digamma^{k}x\right)  .\text{ }%
\]
Now when the initial value $x$ is distributed due to a measure $\nu$ on $I,$
the above $p_{i}^{\left(  n\right)  }$ is given as%

\[
p_{i}^{\left(  n\right)  }\equiv\frac{1}{n+1}\int_{I}\sum_{k=m}^{m+n}1_{A_{i}%
}\left(  \digamma^{k}x\right)  d\nu.\text{ }%
\]
The joint distribution $\left(  p_{ij}^{\left(  n,n+1\right)  }\right)  $
between the time $n$ and $n+1$ is defined by%

\[
p_{ij}^{\left(  n,n+1\right)  }\equiv\frac{1}{n+1}\sum_{k=m}^{m+n}1_{A_{i}%
}\left(  \digamma^{k}x\right)  1_{A_{j}}\left(  \digamma^{k+1}x\right)  \text{
}%
\]
or%

\[
p_{ij}^{\left(  n,n+1\right)  }\equiv\frac{1}{n+1}\int_{I}\sum_{k=m}%
^{m+n}1_{A_{i}}\left(  \digamma^{k}x\right)  1_{A_{j}}\left(  \digamma
^{k+1}x\right)  d\nu.\text{ }%
\]
Then the channel $\Lambda_{n}^{\ast}$ at $n$ is determined by%

\[
\Lambda_{n}^{\ast}\equiv\left(  \frac{p_{ij}^{\left(  n,n+1\right)  }}%
{p_{i}^{\left(  n\right)  }}\right)  \Longrightarrow\varphi^{\left(
n+1\right)  }=\Lambda_{n}^{\ast}\varphi^{\left(  n\right)  },\text{ }%
\]
and the entropic chaos degree is given by the definition 3.1;
\begin{equation}
D_{A}\left(  x;F\right)  =D_{A}\left(  p^{\left(  n\right)  };\Lambda
_{n}^{\ast}\right)  =\sum_{i}p_{i}^{\left(  n\right)  }S(\Lambda_{n}^{\ast
}\delta_{i})=\sum_{i,j}p_{ij}^{\left(  n,n+1\right)  }\log\frac{p_{i}^{\left(
n\right)  }}{p_{ij}^{\left(  n,n+1\right)  }}.
\end{equation}

We can judge whether the dynamics causes a chaos or not by the value of D as
the definition 2.2%

\begin{align*}
D  & >0\Longleftrightarrow chaotic\\
D  & =0\Longleftrightarrow stable.
\end{align*}

This chaos degree was applied to several dynamical maps such logistic map,
Baker's transformation and Tinkerbel map, and it could explain their chaotic
characters. This chaos degree has several merits compared with usual measures
such as Lyapunov exponent as explained below.

$\left(  2\right)  $ \emph{Dynamics is given by }$\varphi_{t}=$ $\digamma
_{t}^{\ast}\varphi_{0}\ on$ \emph{a Hilbert space}: Similarly as making a
difference equation for state, the channel $\Lambda_{n}^{\ast}$ at $n$ is
first deduced from $\digamma_{t}^{\ast}$, which should satisfy $\varphi
^{\left(  n+1\right)  }=\Lambda_{n}^{\ast}\varphi^{\left(  n\right)  }.$ By
means of this constructed channel ($\mathbf{\alpha}$) we compute the chaos
degree $D$ directly according to the definition 3.2 or ($\mathbf{\beta}$) we
take a proper observable $X$ and put $x_{n}\equiv$ $\varphi^{\left(  n\right)
}(X),$ then go back to the algorithm (1).

Note that the chaos degree $D$ does depend on a partition $A$ taken, which is
somehow different from usual degree of chaos (cf., dynamical entropy
\cite{AOW,Ben,AF,KOW}). This is a key point of our understanding of chaos,
which will be discussed in the next section.

\subsection{Logistic Map}

Let us explain how the entropy chaos degree (ECD) well describes to the
chaotic behaivor of logistic map.

The logistic map is defined by%

\[
x_{n+1}=ax_{n}\left(  {1-x_{n}}\right)  ,x_{n}\in\left[  {0,1}\right]  ,0\leq
a\leq4
\]
The solution of this equation bifurcates as shown in Fig.5.1.%

\begin{center}
\includegraphics[
width=3.5674in
]%
{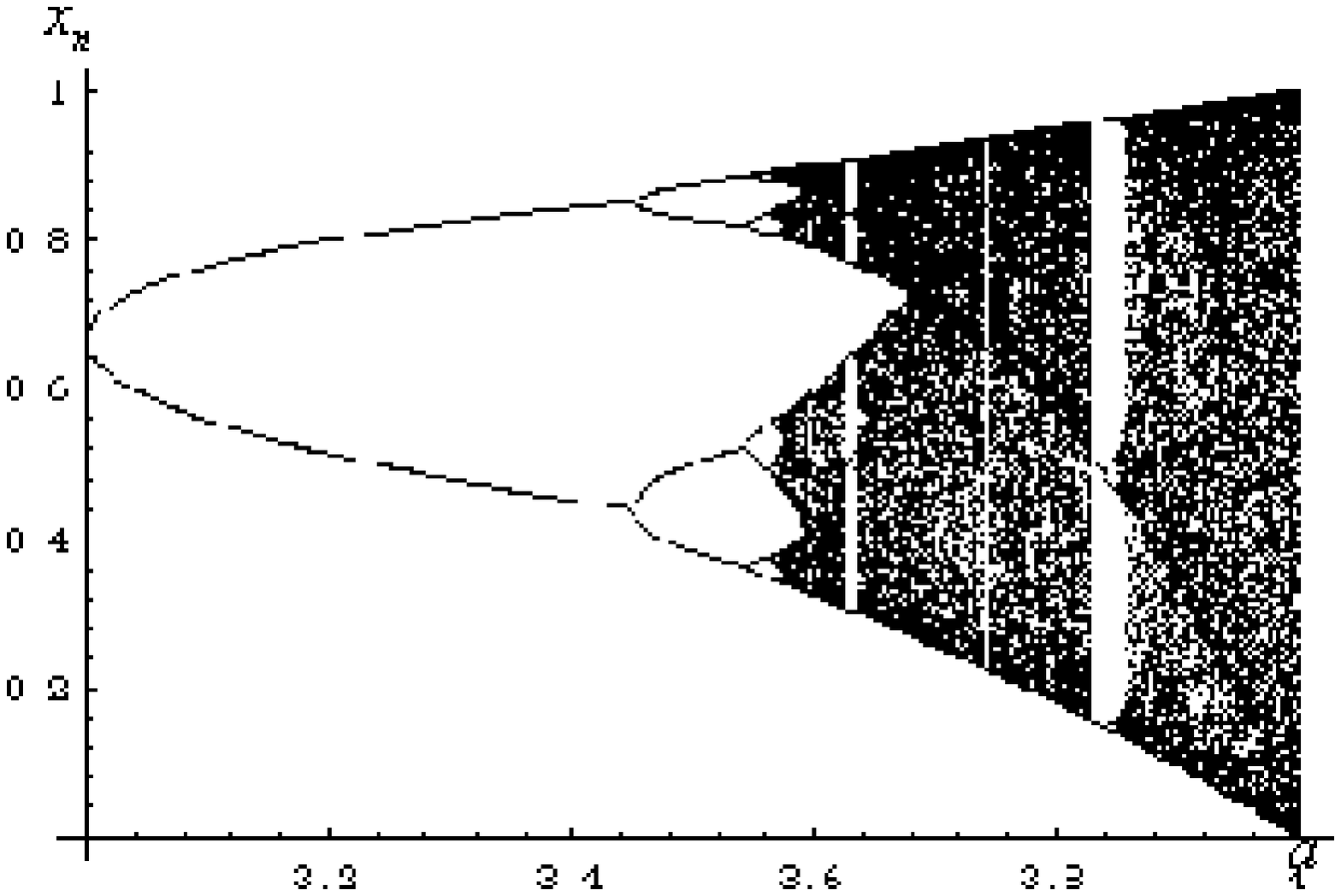}%
\end{center}

\begin{center}
Fig.1 The bifurcation diagram for logistic map
\end{center}

In order to compare ECD with other measure describing chaos, we take Lyapunov
exponent for this comparison and remaind here its definition.

\noindent\textbf{%
$<$%
Lyapunov exponent $\lambda\left(  f\right)  $%
$>$%
}

\begin{itemize}
\item[(1)] Let $f$ be a map on $\mathbf{R}$, and let $x_{0}\in\mathbf{R}$.
Then the Lyapunov exponent $\lambda_{\mathcal{O}}\left(  f\right)  $ for the
orbit $\mathcal{O}\equiv\left\{  {f^{n}\left(  {x_{0}}\right)  ;n=0,1,2,\cdots
}\right\}  $is defined by
\[
\lambda_{\mathcal{O}}\left(  f\right)  ={\lim_{n\to\infty}}\lambda
_{\mathcal{O}}^{\left(  n\right)  }\left(  f\right)  ,\quad\lambda
_{\mathcal{O}}^{\left(  n\right)  }\left(  f\right)  =\frac{1}{n}\log\left|
{\frac{{df^{n}}}{{dx}}\left(  {x_{0}}\right)  }\right|
\]

\item[(2)] \noindent Let $f=\left(  {f_{1},f_{2},\cdots,f_{m}}\right)  $ be a
map on $\mathbf{R}^{m}$, and let $x_{0}\in\mathbf{R}^{m}$. The Jacobi matrix
$J_{n}=Df^{n}\left(  {r_{0}}\right)  $ at $r_{0}$ is defined by
\end{itemize}%

\[
J_{n}=Df^{n}\left(  {r_{0}}\right)  =\left(
\begin{array}
[c]{lll}%
\frac{\partial f_{1}^{n}}{\partial x_{1}}\left(  r_{0}\right)  & \cdots &
\frac{\partial f_{1}^{n}}{\partial x_{m}}\left(  r_{0}\right) \\
\vdots &  & \vdots\\
\frac{\partial f_{m}^{n}}{\partial x_{1}}\left(  r_{0}\right)  & \cdots &
\frac{\partial f_{m}^{n}}{\partial x_{m}}\left(  r_{0}\right)
\end{array}
\right)  .
\]

\noindent Then, the Lyapunov exponent $\lambda_{\mathcal{O}}\left(  f\right)
$ of $f$ for the orbit $\mathcal{O}\equiv\left\{  {f^{n}\left(  {x_{0}%
}\right)  ;n=0,1,2,\cdots}\right\}  $ is defined by
\[
\lambda_{\mathcal{O}}\left(  f\right)  =\log\tilde{\mu}_{1},\quad\tilde{\mu
}_{k}={\lim_{n\to\infty}}\left(  {\mu_{k}^{n}}\right)  ^{\frac{1}{n}}\left(
{k=1,\cdots,m}\right)  .
\]

\noindent Here, $\mu_{k}^{n}$ is the $k$th largest square root of the $m$
eigenvalues of the matrix $J_{n}J_{n}^{T}$.%

\begin{align*}
\lambda_{\mathcal{O}}\left(  f\right)   & >0\Rightarrow\text{Orbit
}\mathcal{O}\text{ is chaotic.}\\
\lambda_{\mathcal{O}}\left(  f\right)   & \le0\Rightarrow\text{Orbit
}\mathcal{O}\text{ is stable.}%
\end{align*}

\noindent

The properties of the logistic map depend on the parameter $a$. If we take a
particular constant $a$, for example, $a=3.71$, then the Lyapunov exponent and
the entropic chaos degree are positive, the trajectory is very sensitive to
the initial value and one has the chaotic behavior.%

\begin{center}
\includegraphics[
width=4.1442in
]%
{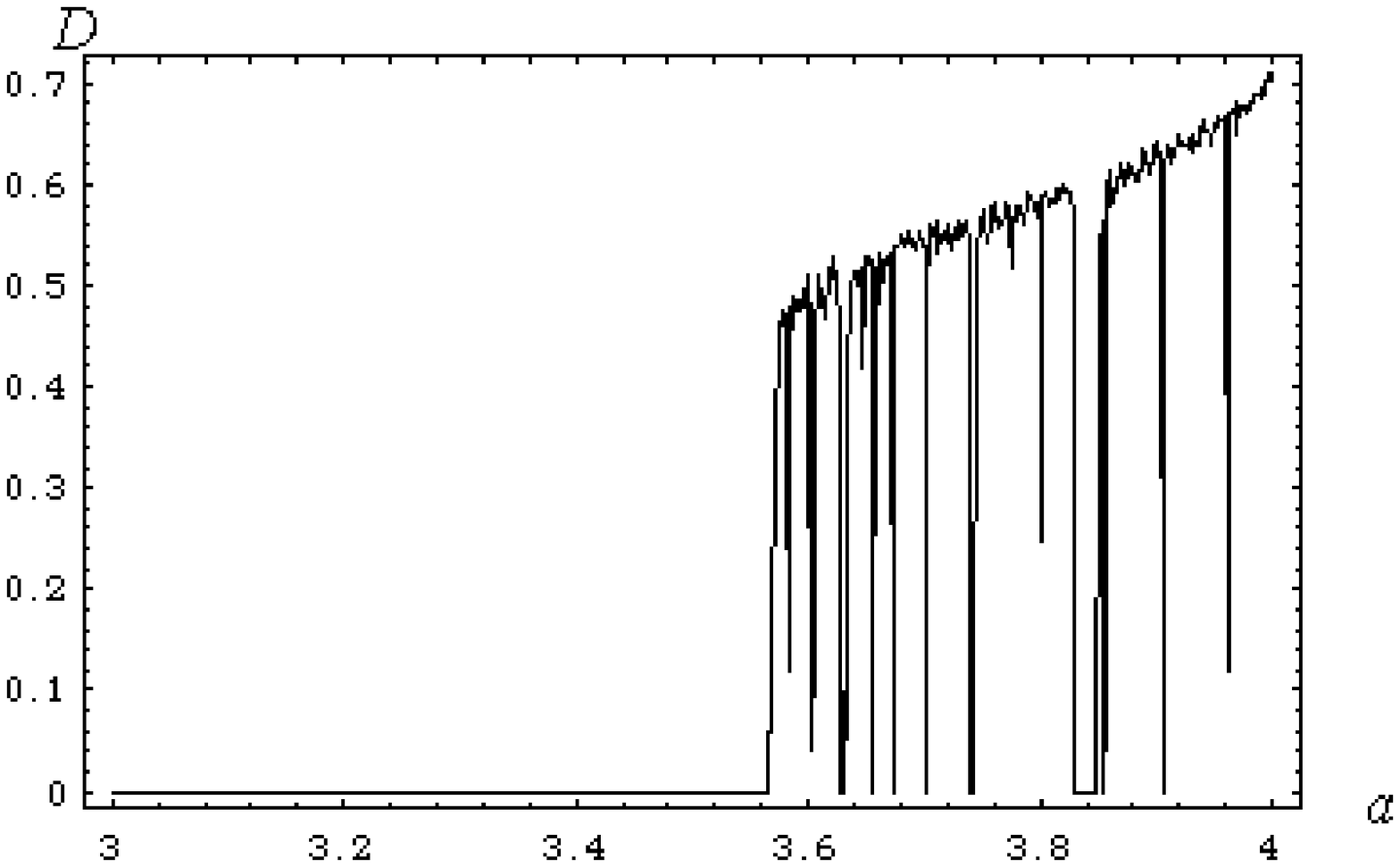}%
\end{center}

\begin{center}
Fig.2. Chaos degree for logistic map
\end{center}%

\begin{center}
\includegraphics[
width=4.0603in
]%
{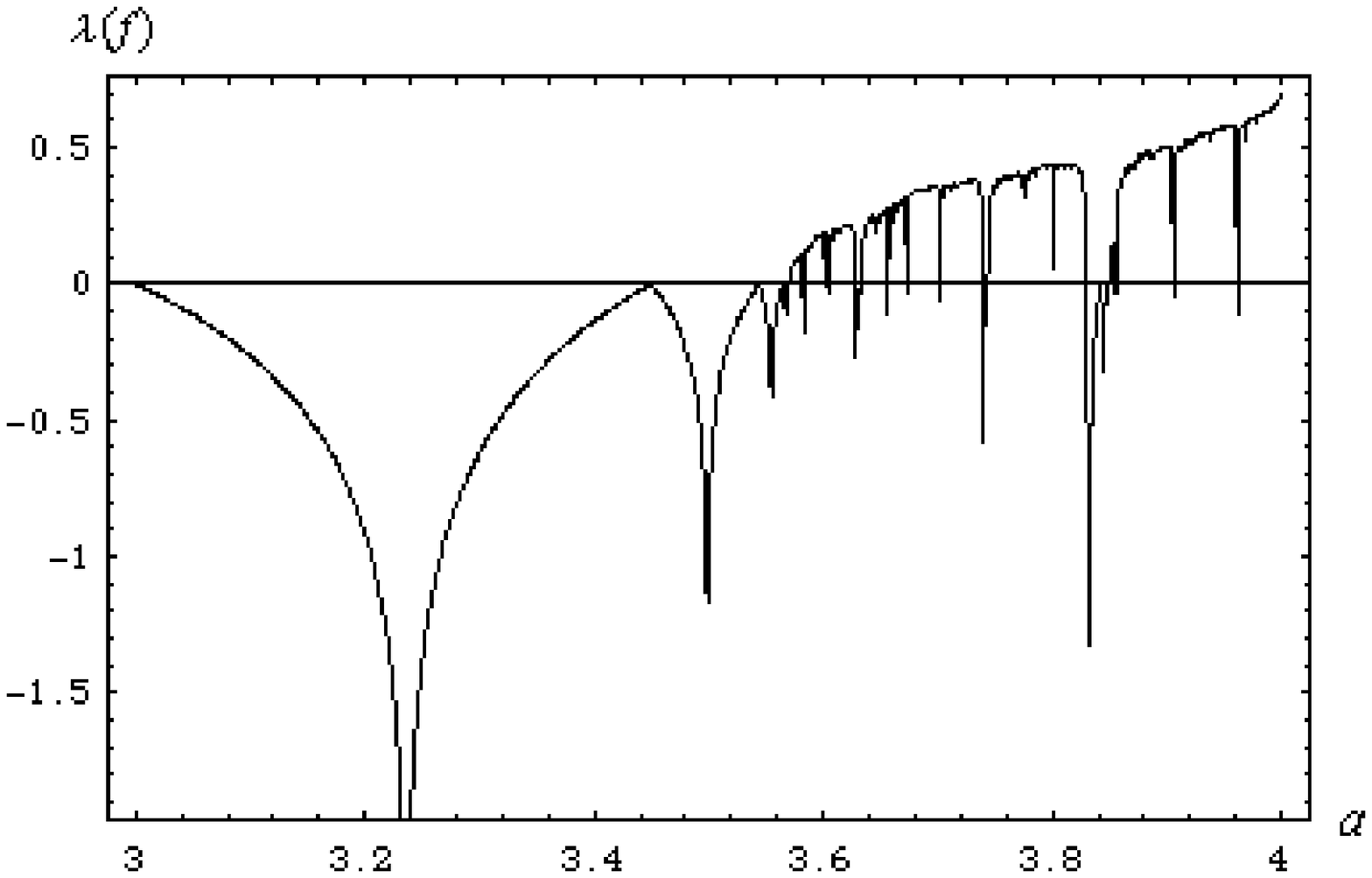}%
\end{center}

\begin{center}
\bigskip Fig.3. Lyapunov exponent for logistic map
\end{center}

From the above example and some other maps (see the paper \cite{IOS}),
Lyapunov exponent and the entropic chaos degree have clear correspondence, but
the ECD can resolve some inconvenient properties of the Lyapunov exponent as follows:

\begin{itemize}
\item[(1)] Lyapunov exponent takes negative value and sometimes $-\infty$, but
the ECD is always positive for any $a\geq0$.

\item[(2)] It is difficult to compute the Lyapunov exponent for some maps like
Tinkerbell map $f$ because it is difficult to compute $f^{n}$ for large $n$.
On the other hand, the ECD of $f\ $is easily computed.

\item[(3)] Generally, the algorithm for the ECD is much easier than that for
the Lyapunov exponent.
\end{itemize}

\section{New Description of Chaos}

First of all we examine carefully when we say that a certain dynamics produces
a chaos. Let us take the logistic map as an example. The original differential
equation of the logistic map is%

\begin{equation}
\frac{dx}{dt}=ax(1-x),0\leq a\leq4
\end{equation}
with initial value $x_{0}$ in $[0,1].$ This equation can be easily solved
analytically, whose solution (orbit) does not have any chaotic behavior.
However once we make the equation above discrete such as%

\begin{equation}
x_{n+1}=ax_{n}(1-x_{n}),0\leq a\leq4.
\end{equation}
This difference equation produces a chaos. Taking the discrete time is
necessary not only to make a chaos but also to observe the orbits drawn by the
dynamics. Similarly as quantum mechanics, it is not possible for human being
to understand any object without observing it, for which it will not be
possible to trace a orbit continuously in time.

Now let us think about finite partition $A$=$\left\{  A_{k};k=1,\cdots
,N\right\}  $of a proper set $I\equiv\left[  a,b\right]  ^{\mathbf{N}}%
\subset\mathbf{R}^{\mathbf{N}}$ and equi-partition $B^{e}=$ $\left\{
B_{k}^{e};k=1,\cdots,N\right\}  $ of $I$. Here ''equi'' means that all
elements $B_{k}^{e}$ are identical. We denote the set of all partitions by
$\mathcal{P}$ and the set of all equi-partitions by $\mathcal{P}^{e}.$ In the
section 3, we specify a special partition, in particular, an equi-partition
for computer experiment calculating the ECD. Such a partition enables to
observe the orbit of a given dynamics, and moreover it provides a criterion
for observing chaos. There exist several reports saying that one can observe
chaos in nature, which are very much related to how one observes the
phenomena, for instance, scale, direction, aspect. It has been difficult to
find a satisfactory theory (mathematics) to explain such chaotic phenomena. In
the difference equation () we take some time interval $\tau$ between $n$ and
$n+1,$ if we take $\tau\rightarrow0$, then we have a complete different
dynamics. If we take coarse graining to the orbit of $x_{t}$ in () for time
during $\tau;$ $x_{n}\equiv\frac{1}{\tau}\int_{(n-1)\tau}^{n\tau}x_{t}dt,$ we
again have a very different dynamics. Moreover it is important for
mathematical consistency to take the limits $n\rightarrow\infty$ or $N$ (the
number of equi-partitions)$\rightarrow\infty$ , i.e., making the partition
finer and finer, and consider the limits of some quantities as describing
chaos, so that mathematical terminologies such as ''lim'', ''sup'', ''inf''
are very often used to define such quantities. In this paper we take the
opposite position, that is, any observation will be unrelated or even
contradicted to such limits. Observation of chaos is a result due to taking
suitable scales of, for example, time, distance or domain, and it will not be
possible in the limiting cases.

\emph{We claim in this paper that most of chaos are scale-dependent phenomena,
so the definition of a degree measuring chaos should dependes on certain
scales taken.}

Taking into cosideration of this view we modify the definitions of the chaos
degree given in the previous sections as below.

Going back to a triple $(\mathcal{A},\mathfrak{S},\alpha\left(  G\right)  )$
considered in Section 2 and we use this triple both for an input and an output
systems. Let a dynamics be described by a mapping $\Gamma_{t}$ with a
parameter $t\in G$ from $\mathfrak{S}$ to $\mathfrak{S}$ and let an
observation be described by a mapping $\mathcal{O}$ from $(\mathcal{A}%
,\mathfrak{S},\alpha\left(  G\right)  )$ to a triple $(\mathcal{B}%
,\mathfrak{T},\beta\left(  G\right)  ).$ The triple $(\mathcal{B}%
,\mathfrak{T},\beta\left(  G\right)  )$ might be same as the original one or
its subsystem and the obserevation map $\mathcal{O}$ may contains several
different types of observations, that is, it can be decomposed as
$\mathcal{O=O}_{m}\mathcal{\cdots}\mathcal{O}_{1}.$Let us list some examples
of observations$.$

For a given dynamics $\frac{d\varphi}{dt}=F\left(  \varphi_{t}\right)  ,$
equivalently, $\varphi_{t}=\Gamma_{t}^{\ast}\varphi,$ one can take several observations.

\begin{example}
Time Scaling (Discretizing): $\mathcal{O}_{\tau}:$ t$\rightarrow n,$
$\frac{d\varphi}{dt}\left(  t\right)  \rightarrow\varphi_{n+1},$ so that
$\frac{d\varphi}{dt}=F\left(  \varphi_{t}\right)  \Rightarrow\varphi
_{n+1}=F\left(  \varphi_{t}\right)  $ and $\varphi_{t}=\Gamma_{t}^{\ast
}\varphi\Rightarrow\varphi_{n}=\Gamma_{n}^{\ast}\varphi.$ Here $\tau$ is a
unit time needed for the observation.
\end{example}

\begin{example}
Size Scaling (Conditional Expectation, Partition): Let $(\mathcal{B}%
,\mathfrak{T},\beta\left(  G\right)  )$ be a subsystem of $(\mathcal{A}%
,\mathfrak{S},\alpha\left(  G\right)  ),$ both of which have a certain
algebraic structure such as C*-algebra or von Neumann algebra. As an example,
the subsystem $(\mathcal{B},\mathfrak{T},\beta\left(  G\right)  )$ has abelian
structure describing a macroscopic world which is a subsystem of a non-abelian
(non-commutative) system $(\mathcal{A},\mathfrak{S},\alpha\left(  G\right)  )$
describing $a$ micro-world. A mapping $\mathcal{O}_{C}$ preserving norm (when
it is properly defined) from $\mathcal{A}$ to $\mathcal{B}$ is, in some cases,
called a conditional expectation. A typical example of this conditional
expectation is according to a projection valued measure $\left\{  P_{k};\text{
}P_{k}P_{j}=P_{k}\delta_{kj}=P_{k}^{\ast}\delta_{kj}\geqq0,\text{ }\sum
_{k}P_{k}=I\right.  \left.  {}\right\}  $ associated with quantum measurement
(von Neumann measurement) such that $\mathcal{O}_{C}\left(  \rho\right)
=\sum_{k}P_{k}\rho P_{k}$ for any quantum state (density operator) $\rho.$
When $\mathcal{B}$ is a von Neumann algebra generated by $\left\{
P_{k}\right\}  ,$ it is an abelian algebra isometrically isomorphic to
$L^{\infty}\left(  \Omega\right)  $ with a certain Hausdorff space $\Omega,$
so that in this case $\mathcal{O}_{C}$ sends a general state $\varphi$ to a
probability measure (or distribution) $p$. Similar example of $\mathcal{O}%
_{C}$ is one coming from a certain representation (selection) of a state such
as a Schatten decomposition of $\mathcal{\rho}$ ; $\rho=\mathcal{O}_{R}%
\rho=\sum_{k}\lambda_{k}E_{k}$ by one-dimensional orthgonal projections
$\left\{  E_{k}\right\}  $ associated to the eigenvalues of $\rho$ with
$\sum_{k}E_{k}=I.$ Another important example of the size scaling is due to a
finite partition of an underlining space $\Omega,$ e.g., space of orbit,
defined as $\mathcal{O}_{P}\left(  \Omega\right)  $=$\left\{  P_{k};P_{k}\cap
P_{j}=P_{k}\delta_{kj}(k,j=1,\cdots N),\text{ }\cup_{k=1}^{N}P_{k}%
=\Omega\right\}  .$
\end{example}

We go back to the discussion of th entropic chaos degree. Starting from a
given dynamics $\varphi_{t}=\Gamma_{t}^{\ast}\varphi,$ it becomes $\varphi
_{n}=\Gamma_{n}^{\ast}\varphi$ after handling the operation $\mathcal{O}%
_{\tau}.$ Then by taking proper combinations $\mathcal{O}$ of the size scaling
operations like $\mathcal{O}_{C},$ $\mathcal{O}_{R}\ $and $\mathcal{O}_{P},$
the equation $\varphi_{n}=\Gamma_{n}^{\ast}\varphi$ changes to $\mathcal{O}%
\left(  \varphi_{n}\right)  =\mathcal{O}\left(  \Gamma_{n}^{\ast}%
\varphi\right)  ,$ which will be written by $\mathcal{O}\varphi_{n}%
=\mathcal{O}\Gamma_{n}^{\ast}\mathcal{O}^{-1}\mathcal{O}\varphi$ or
$\varphi_{n}^{\mathcal{O}}=\Gamma_{n}^{\ast\mathcal{O}}\varphi^{\mathcal{O}}.$
Then our entropic chaos degree is redifined as follows:

\begin{definition}
The entropic chaos degree of $\Gamma^{\ast}$ with an initial state $\varphi$
and observation $\mathcal{O}$ is defined by $D^{\mathcal{O}}\left(
\varphi;\Gamma^{\ast}\right)  =\int_{\mathcal{O}\left(  \mathfrak{S}\right)
}S\left(  \Gamma^{\ast\mathcal{O}}\omega^{\mathcal{O}}\right)  d\mu
^{\mathcal{O}},$ where $\mu^{\mathcal{O}}$ is the measure operated by
$\mathcal{O}$ to a extremal decomposition measure of $\varphi$ selected by of
the observation $\mathcal{O}$ (its part $\mathcal{O}_{R})$.
\end{definition}

\begin{definition}
The entropic chaos degree of $\Gamma^{\ast}$ with an initial state $\varphi$
is defined by $D\left(  \varphi;\Gamma^{\ast}\right)  =\inf$ $\left\{
D^{\mathcal{O}}\left(  \varphi;\Gamma^{\ast}\right)  ;\mathcal{O\in
SO}\right\}  ,$ $where$ $\mathcal{SO}$ is a proper set of observations
natually determined by a given dynamics.
\end{definition}

Then one judges whether a given dynamics causes a chaos or not by the
following way.

\begin{definition}
(1) A dynamics $\Gamma^{\ast}$ is chaotic for an initial state $\varphi$ in an
observation $\mathcal{O}$ iff $D^{\mathcal{O}}\left(  \varphi;\Gamma^{\ast
}\right)  >0.$ (2)A dynamics $\Gamma^{\ast}$ is totally chaotic for an initial
state $\varphi$ iff $D\left(  \varphi;\Gamma^{\ast}\right)  >0.$
\end{definition}

In Definition $,$ $\mathcal{SO}$ is determined by a given dynamics and some
conditions attached to the dynamics, for instance, if we start from a
difference equation with a special representation of an initial state, then
$\mathcal{SO}$ excludes $\mathcal{O}_{\tau}$ and $\mathcal{O}_{R}.$

The idea introducing in this paper to understand chaos can be applied not only
to the entropic chaos degree but also to some other degrees such as dynamical
entropy, whose applications and the comparison of several degrees will be
discussed in the forthcoming paper.

In the caase of logistic map, \ $x_{n+1}=ax_{n}(1-x_{n})\equiv F\left(
x_{n}\right)  ,$ we obtain this difference equation by taking the observation
$\mathcal{O}_{\tau}$ and take an observation $\mathcal{O}_{P}$ by
equi-partition of the orbit space $\Omega=\left\{  x_{n}\right\}  $ so as to
define a state (probability distribution). Thus we can compute the entropic
chaos degree as is discussed in Section 3.

It is important to notice here that the chaos degree does depend on the choice
of observations. As an example, we consider a circle map%

\begin{equation}
\theta_{n+1}=f_{\nu}(\theta_{n})=\theta_{n}+\omega\quad(\mathrm{mod}%
\;2\pi),\label{circle}%
\end{equation}

\noindent where $\omega=2\pi v(0<v<1)$. If $v$ is a rational number $N/M$,
then the orbit $\left\{  \theta_{n}\right\}  $ is periodic with the period
$M$. If $v$ is irrational, then the orbit $\left\{  \theta_{n}\right\}  $
densely fills the unit circle for any initial value $\theta_{0}$; namely, it
is a quasiperiodic motion.

We proved in \cite{IKO2} the following theorem.

\begin{theorem}
Let $\mathbf{I}=[0,2\pi]$ be partioned into $L$ disjoint components with equal
length; $\mathbf{I}=B_{1}\cap B_{2}\cap\ldots\cap B_{L}$.

\begin{itemize}
\item[(1)] If $v$ is rational number $N/M$ , then the finite equi-partition
$P=\left\{  B_{k};k=1,\cdots,M\right\}  $ implies$D^{\mathcal{O}}\left(
\theta_{0};f_{\nu}\right)  =0$.

\item[(2)] If $v$ is irrational, then $D^{\mathcal{O}}\left(  \theta
_{0};f_{\nu}\right)  >0$ for any finite partition P=$\left\{  B_{k}\right\}  $.
\end{itemize}
\end{theorem}

Note that our entropic chaos degree shows a chaos to quasiperiodic circle
dynamics by the observation due to a partition of the orbit, which is
different from usual understanding of chaos. However usual belief that
quasiperiodic circle dynamics will not cause a chaos is not at all obvious,
but is realized in a special limiting case as shown in the following proposition.

\begin{proposition}
For the above circle map, if $v$ is irrational, then $D\left(  \theta
_{0};f_{\nu}\right)  =0.$
\end{proposition}

\begin{proof}
Let take an equipartition $P=\left\{  B_{k}\right\}  $ as
\end{proof}%

\begin{equation}
B_{k}\equiv\left\{  x;2\pi\frac{k-1}{l}\leq x<2\pi\frac{k}{l}\pi\right\}
,\quad k=1,2,\ldots,l,\nonumber
\end{equation}
where $l$ is a certain integer and $B_{k+l}=B_{k}.$ When $\nu$ is irrational,
put $\nu_{0}\equiv\left[  l\nu\right]  $ with Gaussian $\left[  \cdot\right]
$ . Then $f_{\nu}(B_{k})$ intersects only two intervals $B_{k+\nu_{0}}$ and
$B_{k+\nu_{0}+1},$ so that denote the ration of the Lebesgue measure of
$f_{\nu}(B_{k})\cap$ $B_{k+\nu_{0}}$ and that of $f_{\nu}(B_{k})\cap
B_{k+\nu_{0}+1}$ by $1-s:s.$ This $s$ is equal to $l\nu-\left[  l\nu\right]  $
and the entropic chaos degree becomes%

\[
D^{P}=-s\log s-\left(  1-s\right)  \log\left(  1-s\right)  .
\]
Take the continued fraction expasion of $\nu$ and denote its j-th approximate
by $\frac{b_{j}}{c_{j}}$ . Then it holds%

\[
\left|  \nu-\frac{b_{j}}{c_{j}}\right|  \leq\frac{1}{c_{j}^{2}}.
\]
For the above equi-partition $B=\left\{  B_{k}\right\}  $ with $l=c_{j},$ we find%

\[
\left|  l\nu-b_{j}\right|  \leq\frac{1}{k}\text{ }%
\]
and
\[
\left[  l\nu\right]  =\left\{
\begin{array}
[c]{c}%
b_{j}\text{ }\left(  \text{when }\nu-\frac{b_{j}}{c_{j}}>0\right) \\
b_{j-1}\left(  \text{when }\nu-\frac{b_{j}}{c_{j}}<0\right)
\end{array}
\right.  .
\]
It implies%

\[
D^{P}\fallingdotseq\frac{\log c_{j}}{c_{j}},
\]
which goes to 0 as $j\rightarrow\infty.$ Hence $D=\inf\left\{  D^{P}%
;P\right\}  =0.\blacksquare$

Such a limiting case will not take place in real observation of natural
objects, so that we claim that chaos is a phenomenon depending on
observations, which results the definition of chaos above.

In the forthcoming paper \cite{KOT}, we will discuss how to reach to chaos
dynamics by starting from general differential dynamics in both classical and
quantum systems. That is, it is demonstrated how we can get to chaos dynamics
by considering observations introduced in this paper, and we calculate the
entropic chaos degrees in each dynamics.

\subsection{Acknowledgment}

The authors thank JSPS and SCAT for finatial supports in this work.

\end{document}